\documentclass[sigconf,screen]{acmart}
\usepackage{multirow} 
\usepackage{listings}
\usepackage{float}
\usepackage{microtype}
\usepackage{balance}
\AtBeginDocument{%
  }

\setcopyright{cc}
\setcctype{by}
\acmDOI{10.1145/3832783.3837554}
\acmYear{2026}
\copyrightyear{2026}
\acmISBN{979-8-4007-2882-2/2026/10}
\acmConference[ASE '26]{Proceedings of the 41st IEEE/ACM International Conference on Automated Software Engineering}{October 12--16, 2026}{Munich, Germany}
\acmBooktitle{Proceedings of the 41st IEEE/ACM International Conference on Automated Software Engineering (ASE '26), October 12--16, 2026, Munich, Germany}
\begin{document}

\title{The CodeInverter Suite: Structure- and Data-Aware Binary
Decompilation with Efficient LLMs}

\author{Peipei Liu}
\affiliation{%
  \institution{Zhongguancun Laboratory}
  \city{Beijing}
  \country{China}
}
\email{liupp@zgclab.edu.cn}

\author{Jian Sun}
\author{Rongkang Sun}
\affiliation{%
  \institution{Zhongguancun Laboratory}
  \city{Beijing}
  \country{China}
}
\email{\{sunjian,shunrk2025\}@zgclab.edu.cn}

\author{Li Chen}
\authornote{Corresponding authors: Li Chen and Zhaoteng Yan.}

\author{Zhaoteng Yan}
\authornotemark[1]

\affiliation{%
  \institution{Zhongguancun Laboratory}
  \city{Beijing}
  \country{China}
}
\email{\{chenli,yanzt\}@zgclab.edu.cn}

\author{Xiaoling Zhang}
\author{Dawei Wang}
\affiliation{%
  \institution{Zhongguancun Laboratory}
  \city{Beijing}
  \country{China}
}

\author{Dapeng Sun}
\author{Peizheng Zhang}
\affiliation{%
  \institution{Zhongguancun Laboratory}
  \city{Beijing}
  \country{China}
}

\author{Dan Li}
\affiliation{%
  \institution{Tsinghua University}
  \city{Beijing}
  \country{China}
}
\email{tolidan@tsinghua.edu.cn}

\renewcommand{\shortauthors}{P. Liu, J. Sun, R. Sun, L. Chen, Z. Yan, X. Zhang, D. Wang, D. Sun, P. Zhang, and D. Li}

\begin{abstract}
Binary decompilation plays a vital role in various cybersecurity and software engineering tasks.
Recently, end-to-end decompilation methods powered by large language models (LLMs) have attracted increasing attention for their ability to generate highly readable source code with minimal human intervention. However, existing LLM-based methods still struggle with reconstructing program structure and logic, achieving accurate data recovery, ensuring data security and privacy, and maintaining computational efficiency.
To address these challenges, we propose the \textbf{CodeInverter Suite}, with three main pieces: (1) the CodeInverter Workflow (CIW) is a novel prompt engineering method that incorporates control flow graphs (CFG) and explicit data mappings to enhance structure reconstruction and data recovery during decompilation. (2) Building upon CIW, we construct the CodeInverter Dataset (CID), a large-scale domain-specific dataset containing 8.69 million samples enriched with CFGs and data mapping information. (3) We develop CodeInverter Models (CIMs), two lightweight LLMs with 1.3B and 6.7B parameters, enabling efficient inference in privacy-sensitive and resource-constrained environments.
Extensive experiments on two benchmark datasets demonstrate that CIW significantly enhances the decompilation performance of various LLMs, with average improvements of 13.07\% in re-executability and 23.94\% in re-compilability. For our proposed decompilation model, CIM-6.7B achieves state-of-the-art performance in terms of re-executability and readability, outperforming existing LLMs—even with over 100× more parameters—by an average of 11.03\% and 6.27\%, respectively.

\end{abstract}




\maketitle




\section{Introduction}
\label{intros}

Decompilation is a process that restores compiled binary code into functionally equivalent, human-readable high-level programming language code (i.e., \textbf{pscode}\footnote{In general, the recovered code is expressed in pseudo-C form. We use the term \textbf{pscode} as an abbreviation throughout the paper.})\cite{Kruegel2004Detectingkernel,Cheng2019Coda,David2020Nero}. It plays a critical role in numerous software security and digital forensics tasks\cite{Egele2008survey4malware,Szekeres2013SoK,zhou2023binary,Bossert2014protocolre}. For example, when identifying software vulnerabilities or analyzing malware, security researchers often first convert binary executables into interpretable high-level representations due to the absence of source code. 

Currently, end-to-end decompilation approaches with LLMs are gaining increasing attention for generating more readable and syntactically consistent pscode with reduced manual intervention and strong potential for rapid advancement. WaDec~\cite{she2024WaDec} first employs an LLM to decompile WebAssembly. LLM4Decompile~\cite{tan2024LLM4Decompile} supports both direct binary decompilation and Ghidra~\cite{Ghidra2025} output refinement, achieving strong results in readability and executability. Nova~\cite{Nan2024Nova} uses hierarchical attention and contrastive learning to enhance accuracy. FAE~\cite{feng2024sc2dec} recompiles decompiled code for in-context learning and better alignment with source code.


\begin{figure}
  \centering
  \includegraphics[height=1.55in,width=3.25in]{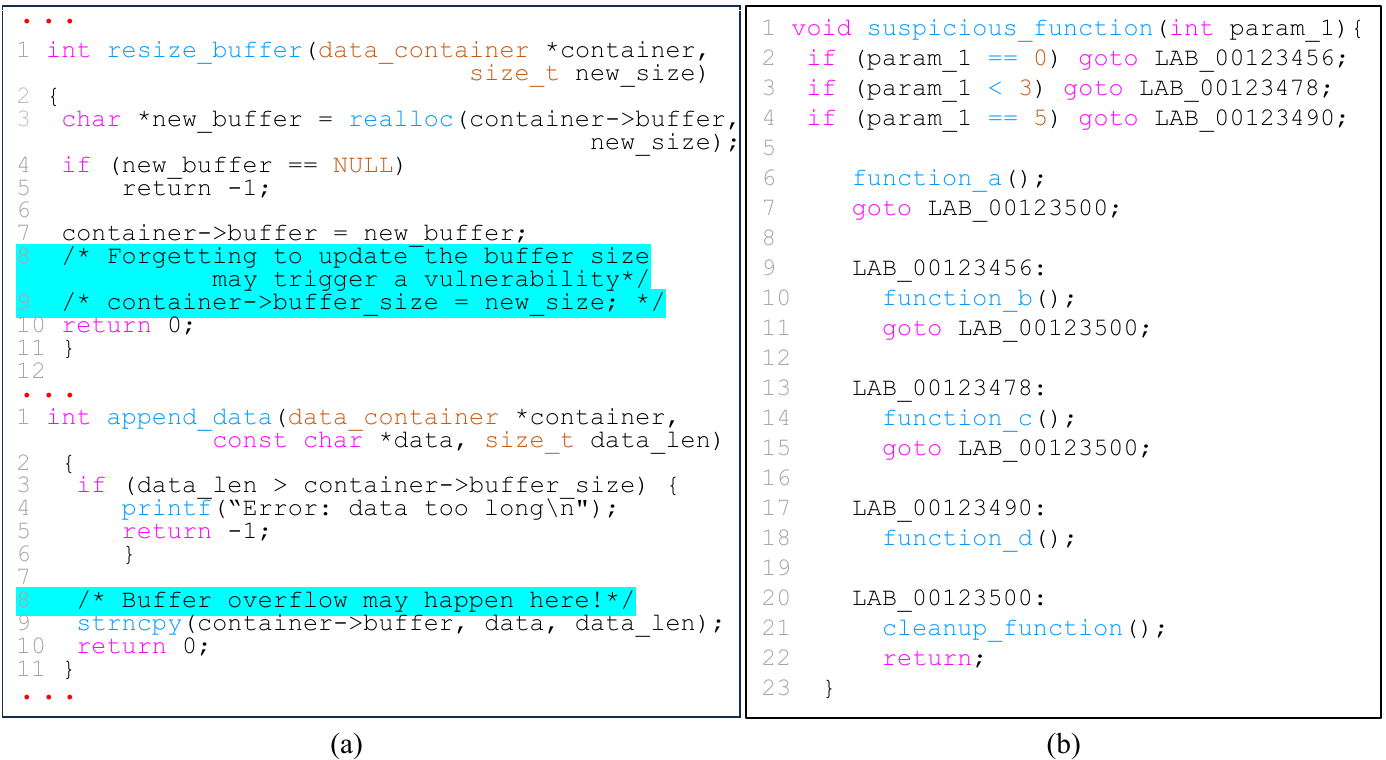}
  \vspace{-1.8ex}
  \caption{(a): The vulnerability reappears in an LLM’s decompilation, (b): Excessive use of \texttt{GOTO} in Ghidra's decompilation}
  \label{introg}
  \vspace{-4ex}
\end{figure}

Despite impressive progress, there still remains practical challenges:
\textbf{(1)} They model assembly instructions as plain text sequences, without explicitly capturing the inherent program structure. This makes it difficult to capture non-local topological constraints in program execution paths, particularly when handling complex control logic (e.g., nested conditional branches, indirect jumps, and loops), cross-block data propagation, and context-dependent relationships, leading to structural fragmentation and logical inconsistencies. 
\textbf{(2)}  
They overlook the data context suggested by memory addresses in assembly instruction sequences, hindering the accurate recovery of data objects and their usage relationships during decompilation. 
For example, in \texttt{mov eax, [rip+0x3fa29]}(\texttt{rip=0x4729}), the model may treat \texttt{[rip+0x3fa29]} merely as a token, ignoring 
the actual data information stored at the address \texttt{.bss:0x44152}.
Furthermore, some LLM-based decompilation works \cite{wong2023DecGPT,hu2024degpt} rely on general-purpose LLMs, including both open-source (e.g., LLaMA~\cite{touvron2023llama}) and commercial models (e.g., GPT\cite{radford2018improving}). 
However, achieving competitive decompilation performance with open-source LLMs requires prohibitively high computational resources and operational costs. For instance, performing inference with a 70B open-source LLM demands at least 4×48GB GPUs (with vLLM + FP16). On the other hand, utilizing third-party commercial models for decompiling binaries may pose significant data security and privacy risks, especially when the binaries contain proprietary algorithms, business-critical logic, or user data. 
{For example, the LLM-generated code contained an identical vulnerability after we used the LLM for CVE analysis (as shown in Fig.\ref{introg}(a)).}

To address the above issues, we propose the \textbf{CodeInverter Suite}, 
consisting of three pieces: the CodeInverter Workflow (\textbf{CIW}), the CodeInverter Dataset (\textbf{CID}), and the CodeInverter Models (\textbf{CIMs}).

The \textbf{CIW} is a novel prompt engineering method with two core ideas according to decompilation principles and properties: (1) CIW integrates the semantic modeling strengths of LLMs with the topological insights of control flow graphs (CFGs) for end-to-end decompilation. By transforming CFGs into sequence-friendly inputs, CIW explicitly incorporates control-flow information into LLM prompts, enabling accurate reconstruction of program structure and control logic. (2) To improve data object recovery, we enhance assembly semantics with an explicit data mapping table extracted from binary data sections such as \texttt{.rodata}, \texttt{.data}, \texttt{.bss}, and \texttt{call stack} layouts. The data mapping reflects relationships between raw memory addresses and their corresponding data, and this mechanism makes implicit data dependencies visible to the model and strengthens data object recovery. 
Guided by the {CIW}, we construct the \textbf{CID}, a large-scale dataset enriched with CFG and data mapping tables annotations, specifically tailored for decompilation. CID provides fundamentally different learning signals from previous work that better reflect deeper semantic modeling in decompilation, shifting the objective from surface-level instruction translation to semantics-aware decompilation.
Trained on the {CID}, we achieve two decompilation LLMs with 1.3B and 6.7B parameters (\textbf{CIM-1.3B/6.7B}) to accommodate privacy-sensitive and resource-constrained local deployment. Compared to prior work, our compact models not only offer several distinct advantages such as deployability, lower computational resource requirements, and complete security control but also significantly improve performance. Experimental results show our models achieve superior performance across major decompilation benchmarks (re-executability, and readability).

In this paper, our contributions can be summarized as follows:

\noindent\textbullet~ We reveal that control-flow structure loss and data-context absence are key factors limiting LLM-based decompilation, and propose CIW to systematically address both issues. To the best of our knowledge, CIW is the first framework to incorporate CFG and data mapping into LLMs, enabling effective knowledge alignment for the decompilation task.


\noindent\textbullet~ We construct the CID dataset, which contains 8.69 million {function-level} assembly–source code pairs for decompilation, each accompanied by semantic labels aligned with control-flow graphs and data-mapping tables for memory addresses. CID provides richer supervision signals that enable the model to better learn inductive biases consistent with decompilation principles.

\noindent\textbullet~ We have trained two lightweight decompilation LLMs CIM-1.3B and CIM-6.7B. Both models can be deployed locally with modest computational resources, ensuring strong data security and privacy.




\noindent\textbullet~ We demonstrate the effectiveness of the CIW and the superiority of the CIMs, and open-source the CIW design process, CID and CIMs to facilitate community research and applications at HuggingFace: \url{https://huggingface.co/CodeInverter}.

\begin{figure*}
  \centering
  \includegraphics[height=3.3in,width=6.8in]{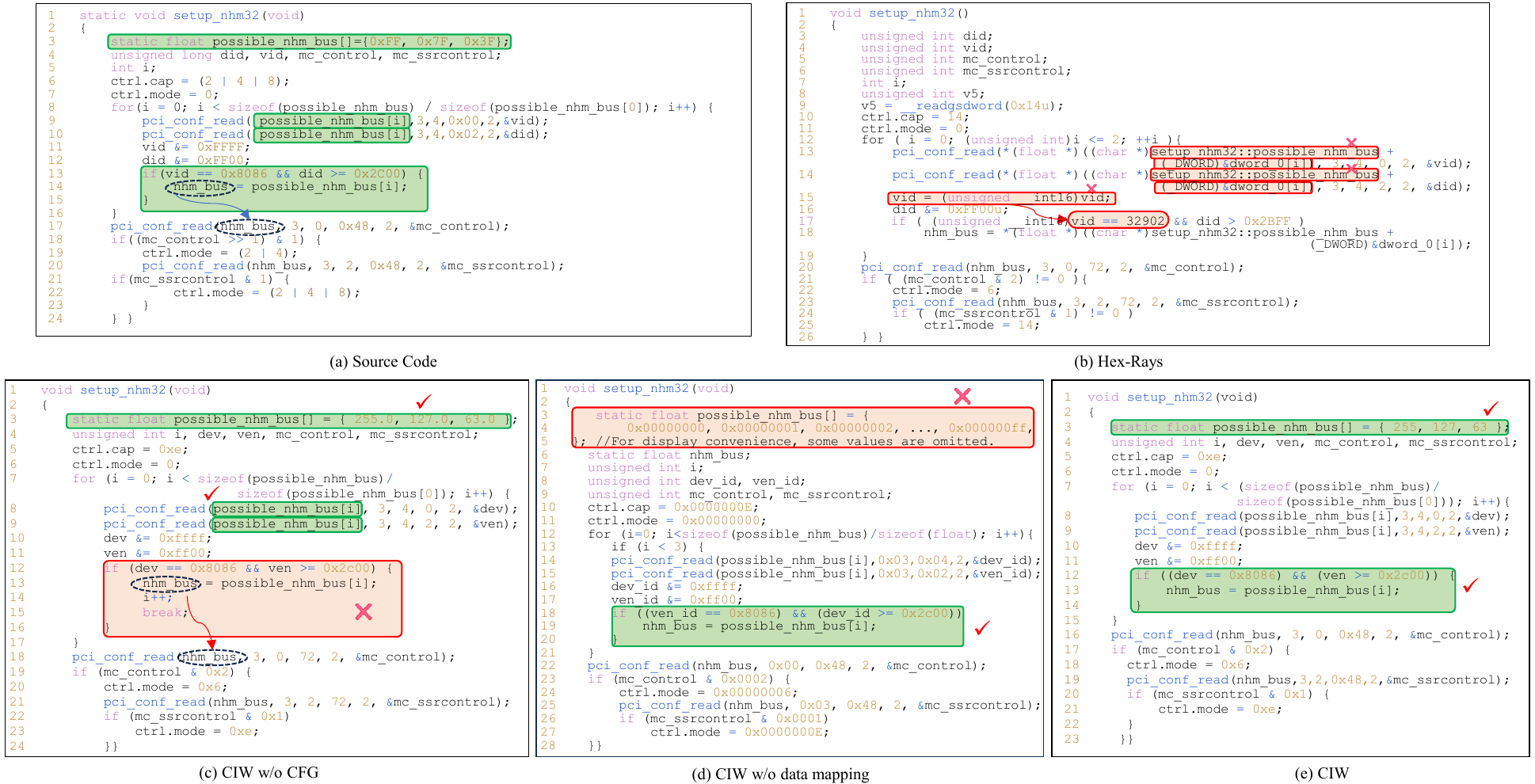}
  \vspace{-1ex}
\caption{Motivating example. Presented are the (a) source code of the case sample alongside the decompilations of (b) Hex-Rays, (c) CIW without CFG on LLM, (d) CIW without data mapping on LLM and (e) CIW on LLM.}
\label{fig:motivation}
  \vspace{-2ex}
\end{figure*}

\section{Background \& Motivation}
\subsection{Control Flow Graph}
CFG is a widely adopted intermediate representation in binary program analysis. It decomposes programs into basic blocks and explicitly models transfer relationships between them (e.g., conditional jumps, loops, and function calls), thereby characterizing all potential execution paths in a program. Compared to linear representations, CFGs provide more accurate topological structures and control dependency information from a static perspective, enabling models to better understand control constraints and dependency relationships between different code blocks.

CFGs have been widely explored in binary analysis. These works support tasks such as binary-code matching~\cite{Yu2020SupplementaryCC}, similarity analysis~\cite{yu2020OrderMatters}, traditional decompiler pipelines~\cite{Cao2022NeurDP}, type inference~\cite{Zhu2024TYGRTI}, and procedure name prediction~\cite{David2020Nero, Gao2021NFRE} by modeling control flow, guiding data-flow analysis, and preserving execution context. However, to the best of our knowledge, few studies have explored using CFGs to assist LLMs in the decompilation process.

\subsection{Decompilation}

To achieve efficient and accurate decompilation, both the research community and industry have proposed a variety of approaches, which can be broadly categorized into following four categories:
{Traditional decompilers},
 such as \texttt{Hex-Rays} \cite{hex-rays2025} and \texttt{Ghidra} \cite{Ghidra2025}, primarily rely on heuristic rules and pattern-matching strategies. 
 Machine/Deep learning-enhanced methods\cite{Cao2022NeurDP,Zhu2024TYGRTI,David2020Nero,Liang2021Neutron,katz2019TraFix,Katz2018recurrent} integrate learning-based architectures to automatically learn mappings from low-level binaries to high-level source code.  
Hybrid methods that combine LLM with traditional decompilers\cite{wong2023DecGPT,hu2024degpt,she2024WaDec} 
first use traditional decompilers to generate pscode,
and then use LLMs to improve and refine the results.
End-to-end decompilation with LLMs\cite{tan2024LLM4Decompile,Nan2024Nova,she2024WaDec,feng2024sc2dec} directly convert assembly code\footnote{{In binary research works, researchers typically use disassembly tools to first convert binary code into assembly code, which then serves as the foundation for downstream task-specific analyses.}} to pscode through prompt engineering or fine-tuning.

Although effective, these methods still have several limitations:
{
\textbf{(I)} Traditional decompilers rely on static program analyses, heuristic rules, 
and pattern-matching strategies for high-level structure recovery, type 
recovery, and pscode generation. Developing and maintaining these 
components across architectures, compilers, and programming languages 
requires substantial engineering effort, limiting scalability. Moreover, the generated pscode often deviates from standard semantic abstractions and coding conventions (e.g., excessive use of \texttt{GOTO} in Fig.\ref{introg}(b)), reducing readability and increasing analysts’ cognitive burden.
}
\textbf{(II)} Seq2Seq-based neural machine translation methods improve readability and extensibility, but limited model capacity often leads to unclear structural boundaries and inadequate modeling of target-language syntax.
\textbf{(III)} Hybrid methods, while improving readability, remain fundamentally constrained by the intermediate representations produced by traditional decompilers. Consequently, they often inherit structural inaccuracies and semantic errors from the first stage, which the subsequent LLM cannot fully rectify, leading to semantically incorrect or incomplete code.

These limitations motivate the exploration of more flexible and scalable paradigms.
In particular, end-to-end decompilation with LLMs has emerged as a promising direction, due to its ability to generate 
syntactically consistent and more readable pscode while offering efficient extensibility and maintainability.

{
In this work, we use IDA as the disassembly tool to facilitate the construction of CFGs and data mapping tables\footnote{{Although IDA is used in our implementation, the process is tool-agnostic and other disassembly tools can also be used to construct CFGs and data mapping tables.}}, and subsequently perform end-to-end decompilation with LLMs.
}
\subsection{Motivations}

Given the proven utility of structural features derived from CFGs in binary analysis, and the increasing adoption of LLMs in decompilation tasks, we aim to explore the integration of CFGs with LLMs to improve the intelligence and effectiveness of the decompilation process. Beyond control flow, we further observe that explicitly mapping memory addresses to their corresponding data contexts can significantly enhance the recovery of data objects, especially in complex binary programs. Such mappings help reveal the relationships between variables, their types, and usage patterns—information that is often obscured or lost during compilation.


To clearly illustrate the motivation behind our work, we present a concrete example that shows the benefits of incorporating both control-flow structures and data-related contextual information into an LLM (DeepSeek-V3).

\begin{figure*}
  \centering
  \includegraphics[height=1.25in,width=5.65in]{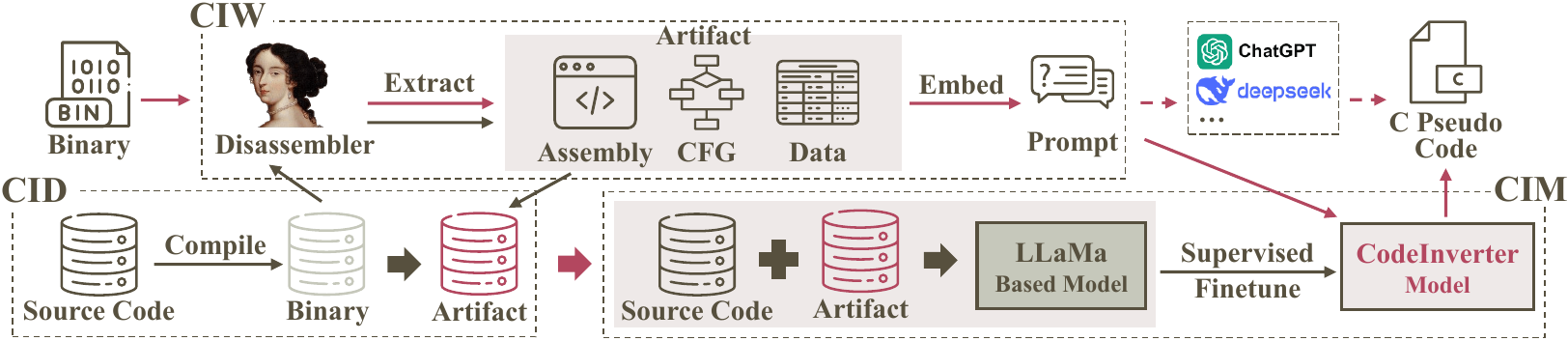}
  \vspace{-1ex}
  \caption{The decompilation workflow with LLMs in this paper (top: inference, bottom: training).}
  \label{overviewfig}
  \vspace{-3ex}
\end{figure*}

The source code of the motivating function \textit{setup\_nhm32} is in Fig.\ref{fig:motivation}(a).
This function’s behavior relies on two critical elements: (1) the initialization of the variable \texttt{possible\_nhm\_bus} (line 3), which is propagated to its caller functions, and (2) the \texttt{if} branch at line 13, where computations influence subsequent control flow.
When compiled with \texttt{-O0} for the x86-32 architecture and decompiled using the Hex-Rays\cite{hex-rays2025}
(Fig.~\ref{fig:motivation}(b)), we observe compound failures: the variable \texttt{possible\_nhm\_bus} remains uninitialized (lines 13-14), and \texttt{vid} is incorrectly assigned (line 15).
These issues jointly corrupt the subsequent branch execution, illustrating the interdependence between value recovery and control-flow reconstruction—errors in one domain can propagate and compromise the other.

These limitations motivate the design of CIW.
The data-mapping-only variant accurately recovers constants (Fig.~\ref{fig:motivation}(c)) but produces incoherent control structures (e.g., lines 14–15). In contrast, the CFG-only version (Fig.~\ref{fig:motivation}(d)) preserves the logical control flow but severely misrecovers values—for instance, expanding the variable \texttt{possible\_nhm\_bus} to 256 data entries, whereas the source defines only 3.
These errors cascade: incorrect values distort the behavior of caller functions, and flawed control flow disrupts value propagation paths.
Only the full CIW (Fig.~\ref{fig:motivation}(e)), which integrates both components, achieves functionally correct decompilation. While minor representational differences remain, the synergy of CFG-guided structural accuracy and precise value recovery resolves the interdependence problem observed in flat text prompt.
This demonstrates that reliable decompilation requires the joint validation of macro-level control structures and micro-level data values—a capability uniquely enabled by our work.






\section{Design of CodeInverter Suite}
\label{method}





\subsection{The Design of CIW}

Unlike the input formulation adopted by LLM4Decompile\cite{tan2024LLM4Decompile}—where raw assembly instructions are concatenated with line-feed characters into a flat string [ASM code] (e.g., \texttt{mov eax, cs:cfg\textbackslash n mov cs:aasworld, eax\textbackslash n retn}) and the LLM is prompted with a generic template \textit{``\# This is the assembly code: [ASM code] \# What is the source code?''}—we carefully craft control-flow graph and data mapping and inject  both information into the LLM prompt input. This approach effectively activates the LLM's specific domain knowledge and guides it toward generating more reasonable and accurate decompilation results.

\subsubsection{CFG-augmented Code Structure Reconstruction}
\label{cfg-section}

In decompilation tasks, achieving code structure and control logic consistent with the original source code requires LLMs to effectively analyze and leverage features including branch logic, loop boundaries, and functional hierarchies in assembly instructions. However, raw assembly instructions are presented as linear sequences and lack explicit representations of these features, making it difficult for LLMs to accurately recover the source code. Considering that the CFG can explicitly model a program’s execution logic and provide LLMs with structured topological cues, we propose to integrate CFGs with LLMs to facilitate the code structure reconstruction. 

\begin{figure*}
  \centering
  \includegraphics[height=1.81in,width=6.2in]{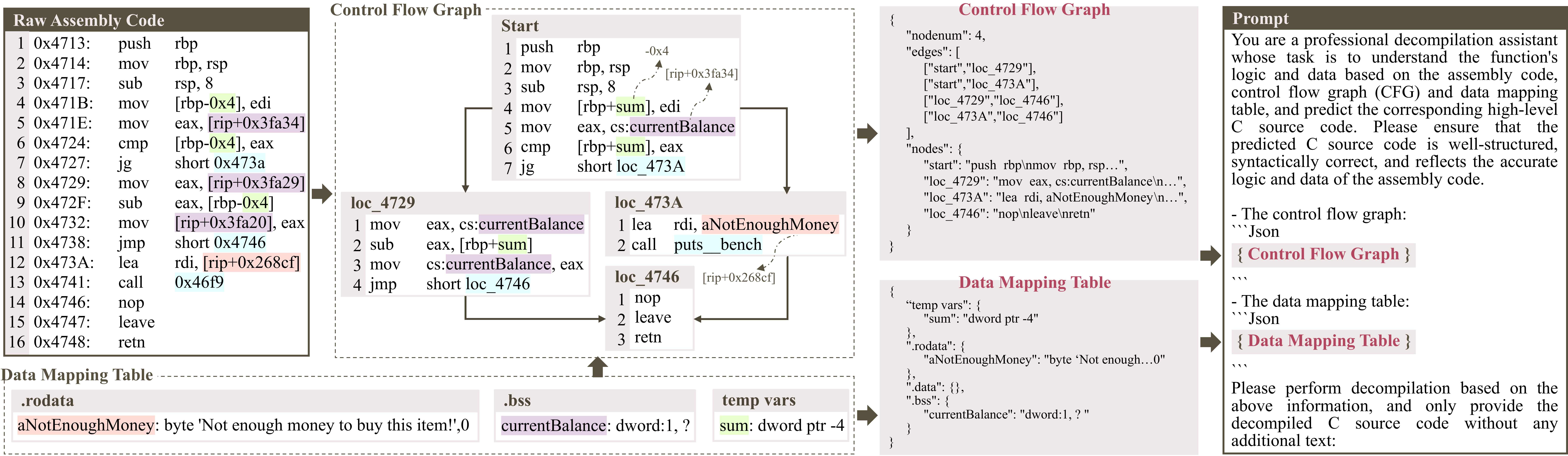}
  \vspace{-1ex}
  \caption{Prompt engineering details of our proposed CIW (the resulting prompt for all LLMs)}
  \label{dataprocessfig}
  \vspace{-3ex}
\end{figure*}


\textbf{CFG Extraction.}  
Based on the assembly code obtained from the input binary, we extract the CFG for target functions using the \texttt{IDA} \cite{hex2014ida}, with assistance from \texttt{objdump} \cite{objdump2025} to improve function boundary identification.  

The process begins with function identification: 
We first use \texttt{IDA} to determine function boundaries. 
When symbol tables are available, we identify function entry points by matching function symbols with their corresponding addresses in the binary.
In cases where \texttt{IDA} fails to identify a target function, we use the starting address derived from \texttt{objdump} to specify the entry point for subsequent control flow analysis.

After that, basic block segmentation is performed within each function recursively, where the instruction sequence is split at each control transfer instruction (e.g., jumps, calls) and their immediate targets. Then, edges between basic blocks are established through direct branch resolution for unconditional and conditional jumps, and through fall-through analysis for sequential execution paths.

\textbf{CFG Representation.} 
The extracted CFG is first represented as a dictionary incorporating well-labeled assembly instruction blocks (as shown in Figure~\ref{dataprocessfig}), and then the dictionary is serialized into a JSON string to meet the input requirements of LLMs.

To improve both machine understanding and processing efficiency, we use a semantically enriched symbolic labeling scheme designed by \cite{hex2014ida}. Each basic block is assigned a unique symbolic label, and the target transfer addresses in control flow instructions are replaced with the corresponding block labels. For instance, the instruction \texttt{jg short 0x473a} is rewritten as \texttt{jg short loc\_473A}, where \texttt{loc\_473A} is the label assigned to the target block. Similarly, function call operands are replaced with function identifiers instead of numeric addresses, such as \texttt{puts\_bench} in \texttt{loc\_473A} block.

Subsequently, we represent the CFG in a dictionary (“Control Flow Graph” in Figure~\ref{dataprocessfig}) consisting of three key fields: \texttt{"nodenum"} (the total number of basic blocks), \texttt{"nodes"} (the instruction content of each block), and \texttt{"edges"} (a list of directed edges between blocks). Each edge is represented as a pair of source and target block labels—for example, \texttt{["start", "loc\_4729"]} indicates a control transfer from the \texttt{start} block to the \texttt{loc\_4729} block. 

\subsubsection{Data Mapping-augmented Data Object Recovery}
\label{dm-section}


During the program compilation, most data objects—including variables and constants—are stored in designated non-code memory regions, with the exception of a small subset of constants that are embedded directly as immediate values within the code region {\texttt{.text}}. These memory regions are organized as follows: the {\texttt{.rodata}} section stores immutable data known at compile time, including string literals, constant arrays, and other \texttt{const}-qualified variables; the {\texttt{.data}} section contains global and static variables that are explicitly initialized with non-zero values; the {\texttt{.bss}} section holds global and static variables that are uninitialized in the source code, which are zero-initialized at runtime; the {\texttt{call stack}} holds temporary variables (e.g., local variables, function arguments) and return addresses, facilitating runtime context preservation and function invocation.

In binary decompilation, distinguishing between the \texttt{.data} and \texttt{.bss} sections aids in inferring variable types and recovering global/static data objects. The \texttt{.rodata} section contains both semantically rich data—such as string literals—and structural elements like jump tables, which are critical for reconstructing constant references and control-related data objects, respectively.
Therefore, leveraging these data sections can improve the completeness and accuracy of source-level data object recovery
compared to prior \texttt{.text}-only methods.
Although ReF Decompile \cite{feng2025ref} considers such sections, it relies on metadata extracted from the source code, making it impractical and unrealistic in real-world scenarios where the source code is unavailable. 
In contrast, we provide a more practical and principled solution by exploiting cues from the binary-level data sections.



\textbf{Data Retrieval.} We build an \texttt{IDA} plugin to retrieve the data associated with each target function. The process begins by identifying relevant data sections (e.g., \texttt{.rodata}, \texttt{.bss}) through parsing the section headers. These sections define the scope of static memory regions that may contain data used by the program.

After identifying these sections, we extract their full contents to form a pool of candidate data items for further analysis. We then perform cross-reference analysis to determine which data items are validly referenced in the assembly instructions. Data items that are explicitly referenced within the address range of the target function are retained as directly associated.
For data items without explicit references, we additionally include those that are in close proximity to already retained items, under the assumption that such proximity may indicate memory adjacency (e.g., array elements). This heuristic helps reduce the risk of omitting relevant data that may be implicitly referenced.


Additionally, we identify temporary stack variables by analyzing stack register addressing and offset values,
as they are crucial for recovering the types and semantics of underlying data objects.

\textbf{Data Mapping Table.}
With the retrieved data, we subsequently replace the raw address identifiers with semantical symbol information available in the binary following \cite{hex2014ida}, such as \texttt{[rip+0x3fa34]} (\texttt{rip=0x471E})\texttt{$\rightarrow$ .bss:0x44152$\rightarrow$ currentBalance} and \texttt{-0x4$\rightarrow$ sum} (Figure~\ref{dataprocessfig}). As a result, we obtain a data mapping table corresponds to the target function, assigning meaningful data items to the memory addresses in the assembly instructions.

As shown in Figure~\ref{dataprocessfig}, each data item includes its size and corresponding value.  Specifically, data sizes are denoted using standard types: \texttt{byte}, \texttt{word}, \texttt{dword} (double word), and \texttt{qword} (quad word); values are primarily represented in hexadecimal format, except when strings are detected-
such as \textit{“Not enough money to buy this item!”}. Strings are identified by scanning for sequences of printable ASCII or Unicode characters terminated by null bytes. Based on contextual and section attributes, appropriate string types (e.g., C-style or Pascal-style) are then assigned. Uninitialized items in the \texttt{.bss} section are denoted with a {“?”}. To concisely represent consecutive uninitialized items of the same size in the \texttt{.bss} section—e.g., array-like structures—a count is appended to the data size. For example, ``\texttt{dword:1,?}'' indicates a single uninitialized \texttt{dword}-sized value. For stack variables, only the offset and data size are recorded, such as \texttt{temp vars} in Figure~\ref{dataprocessfig}.

Similar to the representation of the CFG, the data mapping table is structured as a dictionary and then serialized into a JSON string, allowing LLMs to effectively interpret and utilize the information.





\subsubsection{Decompilation-Oriented Prompt Design}
Through continuous testing and analysis, we have developed CIW, a highly structured prompt paradigm specifically designed for decompilation. By establishing the precise role of a “binary-aware decompilation expert”, it deeply integrates structural constraints from control flow graphs and semantic information from data mapping, effectively triggering the LLM's capacity for deep reasoning and logical deduction. 
The design and test process of CIW is presented in the HuggingFace, and its content detail is shown in the Fig. \ref{dataprocessfig}.

CIW not only enhances the semantic alignment capability of the LLM, leading to more functionally accurate and semantically consistent outputs, but also repositions LLM-based decompilation work from “prompt engineering" to “structural prior modeling". This establishes CIW as a bridge connecting traditional program analysis with LLMs.

\subsection{CID Dataset Construction}
\label{Dataset Construction}


We construct the CID dataset based on the ExeBench training dataset\cite{Jordi2022ExeBench}, a function-level source code dataset that pairs real-world C functions from GitHub with input-output (IO) examples to enable executable supervision. Starting with 1.2 million C functions from ExeBench training dataset, we generate our dataset with 32-bit and 64-bit through following steps (as illustrated in the bottom left of Figure \ref{overviewfig}), where different bits are for model generalization:

\noindent \textbf{Step1: Resolve Data Type Conflicts.} To generate 32-bit executables of C functions, we address data type conflicts between 32-bit standard libraries and Exebench-generated .c files. For instance, we replace the 64-bit-specific definition ``\texttt{typedef unsigned long size\_t;}'' in Exebench with ``\texttt{typedef unsigned int size\_t;}'' to comply with the 32-bit program standards.

\noindent \textbf{Step2: Compile.} The .c files containing the source code of the target functions are compiled using GCC into executables with four different optimization levels (O0, O1, O2, and O3). After compilation, the executables are stripped to remove debugging information for further processing.



\noindent \textbf{Step3: Generate Disassembly Artifact.} 
We disassemble the binary executables using \texttt{IDA}, and extract the corresponding CFGs and data mapping tables based on the processing techniques of CIW (i.e., Section \ref{cfg-section} and Section \ref{dm-section}).

\noindent \textbf{Step4: Data Representation.} Following the CIW paradigm, we integrate CFG and data mapping table to construct each training sample.

\subsection{CIM Training}
\label{modeltraining}

We adopt a sequence-to-sequence framework to train our decompilation
LLM, which maps assembly instruction sequences to corresponding source code sequences. We initialize the model with checkpoints from LLM4Decompile \cite{tan2024LLM4Decompile}, which is based on the LLaMA architecture \cite{touvron2023llama}. The training objective follows the standard approach used in neural machine translation, which is well-suited for decompilation tasks due to their sequential generation nature: given an input sequence, the model learns to predict the corresponding output sequence token by token.


\textbf{Output Prediction Probability.} Given a CIW input sequence $\mathbf{x}_{\text{in}} = [x_1, \ldots, x_m]$ and a target sequence of source code $\mathbf{x}_{\text{out}} = [x_{m+1}, \ldots, x_n]$, the model predicts the probability of $i$-th token autoregressively using a softmax distribution over the decoder's output logits:
{
  \setlength{\abovedisplayskip}{0.5pt}
  \setlength{\belowdisplayskip}{1pt}
\begin{equation}
    P(x_i \mid x_{1:i-1}; \theta) = \text{softmax}\left(\mathbf{W}_o \mathbf{h}_i^{\text{dec}}\right)_{x_i},
\end{equation}}where $\mathbf{h}_i^{\text{dec}}$ is the decoder's hidden state at position $i$, $\mathbf{W}_o$ is the output projection matrix, and $\theta$ denotes all trainable parameters. During training, we apply teacher forcing by conditioning the prediction of $x_i$ on the ground-truth prefix $x_{1:i-1}$.


\textbf{Loss Function.} The model is trained by minimizing the cross-entropy loss between the predicted token distributions and the ground-truth source code tokens. For a sequence of total length $n$, the loss is defined as:
{
  \setlength{\abovedisplayskip}{0.5pt}
  \setlength{\belowdisplayskip}{1pt}
\begin{equation}
    \mathcal{L}(\theta) = -\sum_{i=m+1}^{n} \log P(x_i \mid x_{1:i-1}; \theta),
\end{equation}}where the summation starts at $i = m+1$ to ensure that only the target source code tokens contribute to the loss. This is equivalent to maximizing the conditional likelihood of the target sequence $\mathbf{x}_{\text{out}}$ given the input sequence $\mathbf{x}_{\text{in}}$ under the model parameters $\theta$.







\section{Experiments}
\label{mainexpes}

We comprehensively evaluate our work against popular decompilers and SOTA decompilation LLMs across three metrics on two test dataset, and further dissect the contribution of each module. 
The experiments are conducted to address following research questions:
\begin{itemize}
    \item RQ1: How effective is our proposed CIW approach?
    \item RQ2: How does CIM compare to existing LLMs and {traditional decompilers} in terms of effectiveness?
    \item RQ3: How effective are the individual components of
CIW?
    \item RQ4: Is data-flow information necessary for achieving
accurate decompilation?
    \item {RQ5: What are the computational costs of using LLMs for binary decompilation?}
\end{itemize}

\subsection{Experimental Setups}
\label{mainexesetups}
\subsubsection{Evaluation Metrics}

Following prior work \cite{tan2024LLM4Decompile,Armengol2024SLaDe,Nan2024Nova}, we adopt following evaluation metrics:





\textbf{\textit{Re-compilation Rate}}: This metric evaluates whether the decompiled code generated by our model can be successfully recompiled into an executable binary without errors. A high Recompilation Rate indicates that the decompiled code is syntactically correct and adheres to the constraints of the target programming language.

\textbf{\textit{Re-executability Rate}}: This metric evaluates the functional correctness of the decompiled code. It measures whether the recompiled binary can execute successfully and produce outputs that accurately match the expected results. A high Re-executability indicates that the decompiled code is semantically and functionally correct, preserving the original program’s logical behavior.


{
The \textbf{\textit{Edit Similarity}} is a key metric for evaluating the \textbf{readability} of decompiled code, and it quantifies the semantic similarity between the decompiled code and original code. Following \cite{Armengol2024SLaDe,Katz2018recurrent,Hosseini2022BTC}, we next define normalized edit similarity ($ES(A,B)$), where higher scores indicate better readability of the decompiled code. Specifically, it is calculated as follows:
{
  \setlength{\abovedisplayskip}{3.5pt}
  \setlength{\belowdisplayskip}{3.5pt}
\begin{equation}
ES(A, B) = 1 - \frac{ED(A,B)}{L_B}
\end{equation}}where $A$ represents the prediction sequence, $B$ represents the true sequence, $L$ represents the length of the sequence, and $ED(A,B)$ represents the editing distance between $A$ and $B$. Editing distance is also called Levenshtein distance, which means the minimum number of operations required to convert a sequence $A$ to $B$. It is usually computed by dynamic programming, and its state transition equation is as follows:
\begin{align}
\scalebox{0.8}{$ED(A_i,B_j) = $}
\begin{cases}
\scalebox{0.7}{$max(L_A, L_B)$}&\scalebox{0.7}{$ min(L_A, L_B) = 0$} \\
\scalebox{0.7}{$min$}
\begin{cases}
\scalebox{0.6}{$ED(A_{i-1},B_j)+1$}\\
\scalebox{0.6}{$ED(A_i,B_{j-1})+1$}\\
\scalebox{0.6}{$ED(A_{i-1},B_{j-1})+1$}\\
\end{cases}
&\scalebox{0.7}{$otherwise$ }  \\
\end{cases}
\end{align}
\vspace{-2.5ex}
}

\begin{table*}[h!]
\caption{Decompilation results on HumanEval with only assembly instruction (without {CIW}). (\%)}
\vspace{-2ex}
\centering
    \resizebox{0.87\textwidth}{!}{
   {
   \fontsize{7pt}{6.1pt}\selectfont
\begin{tabular}{c|ccccccccccc}
\toprule
 \multirow{2}{*}{Metric}&  \multirow{2}{*}{Model}&\multicolumn{5}{c}{HumanEval 32-bit}&\multicolumn{5}{c}{HumanEval 64-bit}\\
\cmidrule(lr){3-7} \cmidrule(lr){8-12}
&  & O0 & O1 & O2 & O3 & AVG & O0 & O1 & O2 & O3 & AVG \\
\hline
\multirow{3}{*}{Re-com} 
    & GPT-4o&85.98 &82.32 &83.54 &78.05 &82.47 &89.02 &77.44 &85.98 &79.27 &82.93 \\
    & Deepseek-V3 &\textbf{94.51} &\textbf{87.80} &\textbf{85.98} &\textbf{89.63} &\textbf{89.48} &\textbf{95.73} &\textbf{88.41} &\textbf{89.02} &\textbf{84.15} &\textbf{89.33} \\
    & Qwen-plus &25.61 &37.80 &41.46 &46.34 &37.80 &73.17 &58.54 &61.69 &62.20 &63.90 \\
\midrule
\multirow{3}{*}{Re-exe}
    & GPT-4o&22.56 &11.59 &13.41 &9.51 &14.27 &33.54 &10.98 &14.02 &9.76 &17.08 \\
    & Deepseek-V3 &\textbf{57.93} &\textbf{29.27} &\textbf{33.54} &\textbf{35.37} &\textbf{39.03} &\textbf{66.46} &\textbf{36.59} &\textbf{37.80} &\textbf{37.80} &\textbf{44.66} \\
    & Qwen-plus &0.00 &3.05 &4.27 &4.88 &3.05 &19.51 &7.93 &5.49 &7.93 &10.22 \\
\midrule
\multirow{3}{*}{ES}
    & GPT-4o&34.92 &31.54 &31.37 &29.88 &31.93 &39.02 &30.02 &31.88 &29.69 &32.65 \\
    & Deepseek-V3 &\textbf{42.65} &\textbf{34.16} &\textbf{34.29} &\textbf{33.17} &\textbf{36.07} &\textbf{44.58} &\textbf{34.14} &\textbf{35.75} &\textbf{33.84} &\textbf{37.08} \\
    & Qwen-plus &10.64 &15.78 &16.42 &15.71 &14.64 &27.44 &22.06 &21.57 &22.29 &23.34 \\
\bottomrule
\end{tabular}}%
}
\vspace{-1ex}
\label{human-eval zero}
\end{table*}

\begin{table*}[h!]
\caption{Comparisons between our CIMs with the baselines on HumanEval. *: Our reproduction using LLM4Decompile’s original settings. \dag: Results are from the original paper. (\%)}
\vspace{-2ex}
\centering
    \resizebox{0.87\textwidth}{!}{
{
   \fontsize{7pt}{6.1pt}\selectfont
\begin{tabular}{c|ccccccccccc}
\toprule
 \multirow{2}{*}{Metric}&  \multirow{2}{*}{Model}&\multicolumn{5}{c}{HumanEval 32-bit}&\multicolumn{5}{c}{HumanEval 64-bit}\\
\cmidrule(lr){3-7} \cmidrule(lr){8-12}
&  & O0 & O1 & O2 & O3 & AVG & O0 & O1 & O2 & O3 & AVG \\
\hline
\multirow{7}{*}{Re-com} 
    &GPT-4o +CIW& 95.73 &  91.46 & \textbf{92.07} & \textbf{93.90}& \textbf{93.29} & \textbf{97.56} & 91.46 & \textbf{94.51} & 84.76 & 92.07 \\
    & Deepseek-V3 +CIW& \textbf{96.34} & \textbf{92.07}& 89.63 & 90.85 & 92.22 & \textbf{97.56} & 87.80 & 85.37 & 78.66 & 87.35 \\
    & Qwen-plus +CIW&90.24 &86.59 &85.98 &91.46 &88.57 &89.63 &87.20 &87.20 &70.12 &83.54 \\
    & Ghidra &21.95 & 30.49 & 25.61 & 28.66 & 26.68 &21.95 & 19.51 & 15.85 & 15.24 & 18.14  \\
    & Hex-rays &29.88 &35.37 &35.37 &34.15 &33.69 &26.22 &21.34 &26.83 &25.61 &25.00  \\
    & LLM4Decompile 1.3B* &9.15 &25.00 &17.68 &18.90 &17.68 &56.10 &58.54 &54.27 &56.10 &56.25 \\
    &LLM4Decompile 6.7B* &7.32 &34.15 &35.98 &35.98 &28.35 &71.95 &80.49 &75.00 &75.00 &75.61 \\
    &FAE\dag &- &- &- &- &- &92.07& \textbf{93.29} & 92.07& \textbf{93.90} & \textbf{92.84} \\
    &CIM-1.3B +CIW &85.98 &87.20 &{90.24} &89.36 &88.26 &90.85 &87.80 &87.80 &86.59 &88.26 \\
    &CIM-6.7B +CIW &89.02 &90.85 &{90.24} &92.07 &90.55 &93.29 &91.46 &93.29 &92.07 &92.53 \\
\midrule
\multirow{7}{*}{Re-exe}
    &GPT-4o +CIW& 32.32 & 28.05 & 24.39 & 24.39 & 27.29 & 45.12 & 29.27 & 23.17 & 19.51 & 29.27 \\
    &Deepseek-V3 +CIW& 54.27 & 40.24 & 34.76 & 34.15 & 40.86 & 71.34 & 45.73 & 45.73 & 42.07 & 51.22\\
    &Qwen-plus +CIW&22.56 &17.07 &15.85 &17.68 &18.29 &26.83 &13.41 &12.20 &12.80 &16.31  \\
    &Ghidra &7.32 & 12.20 & 11.59 & 12.20 & 10.83 &19.51 & 14.63 & 12.20 & 11.59 & 14.48    \\
    &Hex-rays &19.51 &27.44 &27.44 &26.83 &25.31 &23.17 &17.07 &21.95 &20.73 &20.73    \\
    &LLM4Decompile 1.3B* &0.00 &1.22 &0.61 &0.61 &0.61 &25.61 &10.37 &7.32 &9.76 &13.26      \\
    &LLM4Decompile 6.7B* &0.61 &1.22 &0.61 &0.61 &0.76 &39.02 &26.22 &30.49 &28.05 &30.94 \\
    &FAE\dag &- &- &- &- &- &71.95 &53.66& 48.78 &45.73 &55.03 \\
    &CIM-1.3B +CIW &65.24 &34.15 &35.37 &31.71 &41.62 &71.34 &39.63 &42.07 &40.24 &48.32 \\
    &CIM-6.7B +CIW &\textbf{75.61} &\textbf{53.05} &\textbf{53.05} &\textbf{50.00} &\textbf{57.93} &\textbf{80.49} &\textbf{57.93} &\textbf{56.71} &\textbf{53.05} &\textbf{62.05} \\
\midrule
\multirow{7}{*}{ES}
    &GPT-4o +CIW& 40.81 & 35.47 & 36.28 & 35.41 & 36.99 & 44.48 & 36.45 & 36.86 & 34.11 & 37.98 \\
    &Deepseek-V3 +CIW& 45.47 & 36.15 & 36.98 &35.33 &38.48 &\textbf{51.36} &36.33 &37.61 &33.61 &39.73 \\
    &Qwen-plus +CIW&39.71 &34.80 &35.76 &34.37 &36.16 &43.59 &35.75 &35.14 &31.63 &36.53 \\
    &Ghidra &28.51 & 26.76 & 26.85 & 26.44 & 27.14 &28.10 & 26.58 & 26.88 & 24.14 & 26.43 \\
    &Hex-rays &31.23 &30.14 &29.21 &28.68 &29.82 &30.74 &29.23 &28.31 &25.26 &28.39 \\
    &LLM4Decompile 1.3B* &14.05 &13.42 &11.23 &11.08 &12.45 &30.14 &19.89 &18.81 &20.26 &22.27 \\
    &LLM4Decompile 6.7B* &29.03 &20.52 &21.80 &21.15 &23.12 &41.42 &32.07 &32.03 &31.68 &34.30 \\
    &CIM-1.3B +CIW &47.45 &35.88 &37.03 &36.33 &39.17 &47.56 &35.67 &37.64 &37.31 &39.55 \\
    &CIM-6.7B +CIW &\textbf{51.06} &\textbf{40.16} &\textbf{39.96} &\textbf{40.48} &\textbf{42.92} &49.16 &\textbf{40.25} &\textbf{39.54} &\textbf{39.02} &\textbf{41.99} \\
\bottomrule
\end{tabular}}%
}
\vspace{-2ex}
\label{humaneval_results}
\end{table*}

\subsubsection{Baselines}
To demonstrate the effectiveness of \textbf{CIW} and the superiority of \textbf{CIM}, we conduct vertical and horizontal evaluations. The vertical evaluation is designed to highlight the importance of \textbf{CIW} in the LLM-driven decompilation, while the horizontal evaluation compares the performance of our LLMs with baselines.

Based on this setting, we select general-purpose LLMs with strong generalization abilities (e.g., GPT-4o \cite{openai2024gpt4}, DeepSeek-V3 \cite{DeepSeekAI2024DeepSeekV3TR}, and Qwen-plus \cite{bai2023qwen}), along with decompilation-specific LLMs (e.g., LLM4Decompile \cite{tan2024LLM4Decompile} and FAE \cite{feng2024sc2dec}) as baselines. In addition, we select two {traditional decompilers} {Hex-Rays} \cite{hex-rays2025} and {Ghidra} \cite{Ghidra2025} as baselines. 
More details of the LLMs are in Supplementary Content A.1.2.

\subsubsection{Evaluation Datasets}
{
Following prior works
~\cite{tan2024LLM4Decompile,feng2024sc2dec,Nan2024Nova}, we use HumanEval \cite{tan2024LLM4Decompile} and ExeBench test set \cite{Jordi2022ExeBench} as evaluation datasets. HumanEval consists of 164 real-world C functions and ExeBench test set contains 5,000 real-world C functions. 
Furthermore, we extend both test datasets to support 32-bit compilation for generalization testing.
} 

{
Since HumanEval and ExeBench are publicly available benchmarks, we consider potential evaluation bias caused by benchmark exposure. We distinguish two possible sources of such bias.
} 

{
(1) CIM may suffer from training-test contamination if the CID contains same functions from the evaluation datasets. To mitigate this risk, we perform multi-level source code decontamination between CID and the evaluation sets. First, we remove exact duplicate functions by computing hashes over normalized source code. Then, we perform AST-based deduplication. Specifically, we parse each source function into an AST and compute a hash value based on its normalized tree structure, where superficial textual differences such as formatting and comments are removed. Training samples whose AST hashes match those of test samples are excluded.
}

\begin{table*}[h!]
\caption{Decompilation results on ExeBench with only assembly instruction (without {CIW}). (\%)}
\vspace{-2ex}
\centering
   \resizebox{0.87\textwidth}{!}{
   { \fontsize{7pt}{6.1pt}\selectfont
\begin{tabular}{c|ccccccccccc}
\toprule
 \multirow{2}{*}{Metric}&  \multirow{2}{*}{Model}&\multicolumn{5}{c}{ExeBench-32-bit}&\multicolumn{5}{c}{ExeBench-64-bit}\\
\cmidrule(lr){3-7} \cmidrule(lr){8-12}
&  & O0 & O1 & O2 & O3 & AVG & O0 & O1 & O2 & O3 & AVG \\
\hline
\multirow{3}{*}{Re-com} 
    & GPT-4o&55.85 &51.58 &50.82 &49.86 &52.03 &68.89 &50.63 &49.21 &\textbf{49.72} &54.61 \\
    & Deepseek-V3 &\textbf{71.04} &\textbf{61.56} &\textbf{60.34} &\textbf{60.02} &\textbf{63.24} &\textbf{79.00} &\textbf{53.76} &\textbf{49.21} &{48.00} &\textbf{57.49} \\
    & Qwen-plus &43.43 &31.85 &31.55 &32.86 &34.92 &57.85 &42.11 &44.59 &39.82 &46.09 \\
\midrule
\multirow{3}{*}{Re-exe}
    & GPT-4o&12.47 &10.30 &8.21 &8.16 &9.79 &24.54 &10.60 &7.89 &7.95 &12.75 \\
    & Deepseek-V3 &\textbf{32.76} &\textbf{20.68} &\textbf{17.97} &\textbf{17.30} &\textbf{22.18} &\textbf{46.75} &\textbf{17.40} &\textbf{12.83} &\textbf{13.63} &\textbf{22.65} \\
    & Qwen-plus &7.87 &1.63 &1.41 &0.79 &2.93 &17.69 &2.19 &4.05 &3.17 &6.78 \\
\midrule
\multirow{3}{*}{ES}
    & GPT-4o&36.59 &30.59 &29.58 &29.70 &31.62 &41.88 &30.90 &29.62 &29.10 &32.88 \\
    & Deepseek-V3 &\textbf{45.22} &\textbf{35.14} &\textbf{33.21} &\textbf{32.84} &\textbf{36.60} &\textbf{50.30} &\textbf{35.00} &\textbf{32.77} &\textbf{32.25} &\textbf{37.58} \\
    & Qwen-plus &32.52 &23.25 &22.03 &21.53 &24.83 &40.25 &29.44 &28.74 &27.67 &31.53 \\
\bottomrule
\end{tabular}}%
}
\vspace{-1.6ex}
\label{exebench zero}
\end{table*}

\begin{table*}[h!]
\caption{Comparisons between our CIMs with the baselines on ExeBench. \dag: Results are from the original paper. (\%)}
\vspace{-2ex}
\centering
    \resizebox{0.87\textwidth}{!}{
   {
   \fontsize{7pt}{6.1pt}\selectfont
\begin{tabular}{c|ccccccccccc}
\toprule
 \multirow{2}{*}{Metric}&  \multirow{2}{*}{Model}&\multicolumn{5}{c}{ExeBench 32-bit}&\multicolumn{5}{c}{ExeBench 64-bit}\\
\cmidrule(lr){3-7} \cmidrule(lr){8-12}
&  & O0 & O1 & O2 & O3 & AVG & O0 & O1 & O2 & O3 & AVG \\
\hline
\multirow{5}{*}{Re-com} 
    &GPT-4o +CIW&80.07 &\textbf{85.25} &\textbf{84.80} &\textbf{85.13} &\textbf{83.81} &90.54 &88.34 &88.09 &87.28 &88.56 \\
    & Deepseek-V3 +CIW&84.30 &81.30 &80.15 &81.41 &81.79 &\textbf{93.32} &\textbf{89.68} &\textbf{89.98} &\textbf{88.68} &\textbf{90.42} \\
    & Qwen-plus +CIW&71.97 &78.68 &78.66 &80.00 &77.33 &82.84 &82.78 &83.20 &81.41 &82.56 \\
    &Ghidra &68.49 & 62.48 & 56.71 & 56.80 & 61.12 &68.48 & 69.80 & 67.10 & 65.86 & 67.81 \\
    &Hex-rays &65.10 &57.32 &48.57 &44.79 &53.94 &56.96 &57.88 &51.71 &49.96 &54.13 \\
    &CIM-1.3B +CIW &84.14 &73.55 &74.72 &73.28 &76.42 &88.33 &70.76 &70.72 &69.99 &74.95 \\
    &CIM-6.7B +CIW &\textbf{86.87} &73.77 &73.01 &73.52 &76.79 &89.49 &70.57 &70.20 &68.14 &74.60\\
\midrule
\multirow{7}{*}{Re-exe}
    &GPT-4o +CIW&30.87 &30.20 &28.27 &29.08 &29.61 &43.99 &25.61 &23.61 &22.06 &28.82 \\
    &Deepseek-V3 +CIW&43.83 &36.55 &34.26 &34.51 &37.29 &59.16 &36.48 &32.89 &30.86 &39.85\\
    &Qwen-plus +CIW&22.05 &22.54 &21.22 &21.28 &21.77 &32.96 &19.85 &17.33 &16.33 &21.62\\
    &Ghidra &61.48 & \textbf{57.42} & \textbf{44.91} & \textbf{44.45} & \textbf{52.06} &63.79 & \textbf{66.31} & \textbf{57.43} & \textbf{58.53} & \textbf{61.51}\\
    &Hex-rays &57.76 &50.63 &39.04 &37.72 &46.29 &54.67 &52.99 &43.05 &37.34 &47.01\\
    &LLM4Decompile 1.3B\dag &- &- &- &- &- &17.86 &13.62 &13.20 &13.28 &14.49 \\
    &LLM4Decompile 6.7B\dag &- &- &- &- &- &22.89 &16.60 &16.18 &16.25 &17.98\\
    &CIM-1.3B +CIW &56.09 &35.33 &33.94 &32.49 &39.46 &65.20 &36.32 &33.34 &32.55 &41.85 \\
    &CIM-6.7B +CIW &\textbf{64.74} &{42.86} &{40.60} &{40.66} &{47.21} &\textbf{72.13} &{40.42} &{36.57} &{35.45} &{46.14} \\
\midrule
\multirow{5}{*}{ES}
    &GPT-4o +CIW &52.02 &40.53 &38.68 &38.44 &42.42 &46.53 &41.24 &39.85 &40.16 &41.95\\
    &Deepseek-V3 +CIW &54.93 &44.73 &42.97 &42.70 &46.33 &60.95 &44.31 &41.81 &41.20 &47.07\\
    &Qwen-plus +CIW &43.62 &40.39 &39.15 &39.06 &40.56 &50.98 &39.87 &38.59 &37.70 &41.79\\
    &Ghidra &55.45 & 45.91 & 41.96 & 41.30 & 46.16 &50.57 & 46.18 & 43.41 & 42.60 & 45.69\\
    &Hex-rays &27.98 &26.54 &26.19 &26.33 &26.76 &27.08 &27.37 &27.00 &27.03 &27.12\\
    &CIM-1.3B +CIW &66.14 &50.02 &49.03 &48.03&53.31 &67.58 &50.27 &49.21 &48.28 &53.84\\
    &CIM-6.7B +CIW &\textbf{69.94} &\textbf{53.04} &\textbf{51.28} &\textbf{50.68} &\textbf{56.24} &\textbf{67.94} &\textbf{52.72} &\textbf{51.03} &\textbf{50.48} &\textbf{55.54}\\ 
\bottomrule
\end{tabular}}%
}
\vspace{-2ex}
\label{exebench_results}
\end{table*}


{
(2) For general-purpose baseline LLMs, such as GPT-4o, DeepSeek-V3, and Qwen-plus, we cannot fully audit whether their pretraining corpora contain benchmarks. This uncertainty may affect absolute comparisons among independently pretrained LLMs. However, the comparison between CIM and LLM4Decompile provides a more controlled setting for analyzing potential benchmark exposure. CIM is initialized from the LLM4Decompile and is further trained on the decontaminated CID. Therefore, potential benchmark exposure inherited from the same base LLM is controlled in this comparison. After removing overlaps between CID and the evaluation benchmarks, the additional performance gain of CIM over LLM4Decompile can be attributed to the structure- and data-aware training signals introduced by CID and CIW, rather than direct benchmark memorization during fine-tuning.
} 

{
To further reduce the influence of benchmark-reuse bias, we additionally evaluate our CIM on an independent real-world application beyond existing benchmarks, as detailed in evaluation section~\ref{sec:real_world}.
}

\subsubsection{Training and Inference Configuration}
We train both 1.3B and 6.7B parameter LLM for one epoch (15 steps). 
The batch sizes are set to 16 and 12 for the 1.3B and 6.7B models, respectively. We employ the AdamW optimizer with an initial learning rate of 2e-5, and set the maximum sequence length to 4096. Training is conducted on an NVIDIA GPU cluster with 8×H100-80GB, and we utilize ColossalAI for efficient distributed training.

To enhance generation performance, we adopt a hierarchical training strategy, gradually introducing samples from simple to complex based on the number of CFG blocks. During inference and evaluation, we consistently apply greedy decoding. The decompilation process is accelerated using vLLM \cite{kwon2023vllm}, which also ensures deterministic output.
\subsection{Main Results}
The main experimental results are presented in Table~\ref{human-eval zero}--\ref{humaneval_results} for HumanEval, and Table~\ref{exebench zero}--\ref{exebench_results} for ExeBench. 
Tables~\ref{human-eval zero} and \ref{exebench zero} report decompilation performance of general-purpose LLMs using our prompt but only with raw assembly sequences. 
Tables~\ref{humaneval_results} and \ref{exebench_results} present the overall results, where general-purpose LLMs are enhanced with \textbf{CIW}, while decompilation-specific LLMs are either reproduced or directly cited from their original papers, depending on the availability of implementations. 
Results from {traditional decompilers} are also included for comparison.
In the following, we will present some findings and analysis of the results.


\begin{table*}[h!]
\caption{Ablation study for CFG. (\%)}
\vspace{-2ex}
\centering
\resizebox{0.87\textwidth}{!}{%
   {
   \fontsize{7pt}{6.1pt}\selectfont
\begin{tabular}{c|ccccccccccc}
\toprule
 \multirow{2}{*}{Metric}&  \multirow{2}{*}{Model}&\multicolumn{5}{c}{ExeBench 64-bit}&\multicolumn{5}{c}{HumanEval 64-bit}\\
\cmidrule(lr){3-7} \cmidrule(lr){8-12}
&  & O0 & O1 & O2 & O3 & AVG & O0 & O1 & O2 & O3 & AVG \\
\midrule
\multirow{4}{*}{Re-com} 
    &CIM-1.3B +CIW &\textbf{87.34} &\textbf{65.02} &\textbf{62.70} &\textbf{62.35} &\textbf{69.35} &91.77 &88.54 &89.17 &87.90 &89.35\\
    &\ w/o cfg  &87.02 &63.14 &61.44 &61.18 &68.19&90.51 &90.45 &90.45 &89.17 &90.15 \\
    &CIM-6.7B +CIW  &87.23 &62.75 &59.77 &61.81 &67.89 &\textbf{93.04} &\textbf{92.99} &\textbf{94.27} &\textbf{92.99} &\textbf{93.32}\\
    &\ w/o cfg  &85.98 &62.06 &60.29 &60.57 &67.23 &92.41 &91.08 &88.54 &90.45 &90.62\\
\midrule
\multirow{4}{*}{Re-exe}
    &CIM-1.3B +CIW  &61.55 &30.61 &28.32 &25.65 &36.53&72.15 &38.85 &42.04 &40.13 &48.29\\
    &\ w/o cfg  &59.87 &28.96 &26.22 &24.15 &34.80&51.27 &25.48 &26.75 &25.48 &32.25\\
    &CIM-6.7B +CIW  &\textbf{67.72} &\textbf{32.33} &\textbf{28.62} &\textbf{27.95} &\textbf{39.04} &\textbf{80.38} &\textbf{58.60} &\textbf{57.32} &\textbf{53.50} &\textbf{62.45} \\
    &\ w/o cfg  &58.61 &28.75 &25.89 &24.15 &34.35&65.19 &34.39 &41.40 &37.58 &44.64\\
\midrule
\multirow{4}{*}{ES}
    &CIM-1.3B +CIW&67.66 &50.76 &49.29 &48.66 &54.09  &48.05 &36.06 &38.27 &37.92 &40.08   \\
    &\ w/o cfg&66.83 &50.60 &48.53 &47.61 &53.39  &45.28 &33.38 &31.70 &31.14 &35.38  \\
    &CIM-6.7B +CIW   &\textbf{68.65} &\textbf{52.30} &\textbf{50.95} &\textbf{49.85} &\textbf{55.44}&\textbf{49.86} &\textbf{40.97} &\textbf{40.09} &\textbf{39.56} &\textbf{42.62} \\
    &\ w/o cfg &66.69 &50.00 &48.40 &47.44 &53.13&45.59 &34.83 &34.72 &34.88 &37.51 \\
\bottomrule
\end{tabular}}%
}
\vspace{-1ex}
\label{nocfg ours}
\end{table*}



\begin{table*}[h!]
\caption{Ablation study for data mapping table. (\%)}
\vspace{-2ex}
\centering
\resizebox{0.87\textwidth}{!}{%
   {
   \fontsize{7pt}{6.1pt}\selectfont
\begin{tabular}{c|ccccccccccc}
\toprule
 \multirow{2}{*}{Model}&  \multirow{2}{*}{Metric}&\multicolumn{5}{c}{ExeBench 64-bit}&\multicolumn{5}{c}{HumanEval 64-bit}\\
\cmidrule(lr){3-7} \cmidrule(lr){8-12}
&  & O0 & O1 & O2 & O3 & AVG & O0 & O1 & O2 & O3 & AVG \\
\midrule
\multirow{4}{*}{Re-com} 
    &CIM-1.3B +CIW &84.37 &60.67 &58.89 &57.10 &65.26&92.50 &87.80 &92.31 &88.68 &90.32 \\
    &\ w/o DM &80.18 &60.39 &58.33 &58.45 &64.34&\textbf{95.00} &90.24 &92.31 &90.57 &92.03 \\
    &CIM-6.7B +CIW &\textbf{86.86} &\textbf{62.64} &\textbf{60.56} &59.25 &\textbf{67.33}&87.50 &\textbf{92.68} &\textbf{94.87} &\textbf{94.34} &\textbf{92.35} \\
    &\ w/o DM &84.03 &56.74 &59.44 &\textbf{61.39} &65.40&\textbf{95.00} &90.24 &84.62 &84.91 &88.69 \\
\midrule
\multirow{4}{*}{Re-exe}
    &CIM-1.3B +CIW  &55.83 &23.31 &14.72 &13.40 &26.82&67.50 &31.71 &38.46 &33.96 &42.91 \\
    &\ w/o DM  &52.89 &21.35 &14.44 &12.60 &25.32&55.00 &26.83 &33.33 &32.08 &36.81 \\
    &CIM-6.7B +CIW &\textbf{65.69} &\textbf{28.93} &\textbf{18.89} &\textbf{17.96} &\textbf{32.87}&\textbf{75.00} &\textbf{48.78} &\textbf{43.59} &\textbf{43.40} &\textbf{52.69} \\
    &\ w/o DM   &60.93 &22.47 &17.50 &17.43 &29.58&55.00 &41.46 &41.59 &41.51 &44.89\\
\midrule
\multirow{4}{*}{ES}
    &CIM-1.3B +CIW  &68.43 &47.89 &39.43 &38.28 &48.51&42.94 &33.10 &\textbf{36.59} &36.32 &37.24\\
    &\ w/o DM   &66.48 &47.63 &38.80 &38.45 &47.84&45.14 &34.58 &33.12 &35.65 &37.12\\
    &CIM-6.7B +CIW&\textbf{71.61} &\textbf{51.81} &42.98 &\textbf{42.91} &\textbf{52.33} &47.41 &36.40 &35.78 &37.24 &39.21  \\
    &\ w/o DM  &70.38 &48.24 &\textbf{44.18} &42.28 &51.27&\textbf{48.16} &\textbf{37.76} &35.02 &\textbf{37.51} &\textbf{39.61} \\
\bottomrule
\end{tabular}}%
}
\vspace{-1ex}
\label{nodata ours}
\end{table*}

\subsubsection{The Effectiveness of CIW (RQ1)}
By comparing Table \ref{human-eval zero} with Table \ref{humaneval_results} and Table \ref{exebench zero} with Table \ref{exebench_results} through vertical evaluation, we observe that \textbf{CIW} significantly enhances decompilation performance on both HumanEval and ExeBench datasets. Specifically, on the HumanEval dataset, the three models achieve absolute improvements of 15.19\% on Re-com, 9.16\% on Re-exe, and 8.36\% on ES; whereas on the ExeBench dataset, the corresponding improvements reach 32.67\%, 16.98\%, and 10.85\%, respectively. Notably, meanwhile, Qwen-plus attains the largest improvement on Re-com (35.21\%) and ES (17.36\%) in HumanEval, while GPT-4o shows the largest improvement on Re-exe (12.61\%); correspondingly, on the ExeBench dataset, Qwen-plus achieves the maximum enhancement on Re-com (39.44\%) and ES (13.0\%), and GPT-4o obtains the most substantial gain on Re-exe (17.95\%).

Based on the observed performance improvements across different evaluation metrics demonstrated by various models on diverse datasets, we can confidently conclude that \textbf{CIW} plays a significantly positive role in decompilation. The consistent enhancement in performance metrics across multiple testing scenarios further demonstrates its robust generalization capability and strong adaptability to different code structures.


\subsubsection{The Advancement of CIM Compared to other LLMs (RQ2)}
As shown in Tables \ref{humaneval_results} and \ref{exebench_results}, in the horizontal evaluation, our CIM-6.7B LLM achieves SOTA performance on Re-exe and the ES metric among LLM-driven decompilations, 
with average improvements of 11.03\% and 6.27\% (on two datasets), respectively, over DeepSeek-V3 671B—a model with 100× more parameters.
This highlights our model's strength in preserving semantic consistency and functional correctness. Although our Re-com score is slightly lower than that of some general-purpose LLMs, this is primarily because they possess stronger contextual modeling capabilities within more parameters. This allows them to generate syntactically complete code, even if it may not faithfully reflect the original program's logic.

\subsubsection{Comparing CIM with {Traditional Decompilers} (RQ2)}
From Table \ref{humaneval_results} and \ref{exebench_results}, we have the following insights:
(1) On the Re-com metric, our CIMs significantly outperform {traditional decompilers} on both datasets. Specifically, the performance gap is most substantial between CIM-6.7B (91.54\%) and Ghidra (22.41\%) on HumanEval, reaching 69.13\%. In contrast, the gap is smallest between CIM-1.3B (75.69\%) and Ghidra (64.47\%) on Exebench, at 11.22\%. This indicates that LLMs surpass rule-based {traditional decompilers} in dimensions such as "semantic restoration" and "re-compilability", primarily due to their ability to effectively leverage internal knowledge to enhance the completeness of the generated code.
(2) On the Re-exe metric, our CIMs and {traditional decompilers} shows a see-saw relationship. On the HumanEval dataset, even the sub-optimal CIM-1.3B (44.97\%) outperforms the best-performing traditional decompiler, Hex-rays (23.02\%). However, on the Exebench dataset, the best-performing CIM-6.7B (46.68\%) only matches the weakest-performing traditional decompiler, Hex-rays (46.65\%), and is inferior to the best-performing traditional decompiler, Ghidra (56.79\%). This suggests that while current LLMs offer greater robustness, a gap remains in core functionalities compared to developed decompilers. This is reasonable because {traditional decompilers} incorporate more targeted rules and domain knowledge, whereas CIM currently integrates only a few rules. Although the model itself may contain rich professional knowledge, more effective prompting methods are required to trigger and guide this knowledge. 
(3) On the ES metric, the CIMs perform excellently on both datasets, indicating a clear advantage of LLMs in the readability of decompiled output.

\begin{table*}[h!]
\caption{Decompilation results on HumanEval with CIW and DUG. (\%)}
\vspace{-2ex}
\centering
    \resizebox{0.87\textwidth}{!}{
   {
   \fontsize{7pt}{6.1pt}\selectfont
\begin{tabular}{c|ccccccccccc}
\toprule
 \multirow{2}{*}{Metric}&  \multirow{2}{*}{Model}&\multicolumn{5}{c}{HumanEval 32-bit}&\multicolumn{5}{c}{HumanEval 64-bit}\\
\cmidrule(lr){3-7} \cmidrule(lr){8-12}
&  & O0 & O1 & O2 & O3 & AVG & O0 & O1 & O2 & O3 & AVG \\
\hline
\multirow{3}{*}{Re-com} 
& GPT-4o +CIW +DUG &81.71          & 82.93          & 84.15          & 78.05          & 81.71          & 89.63          & 79.27          & 89.03          & 78.66          & 84.15          \\
& Deepseek-V3 +CIW +DUG &\textbf{92.65} & \textbf{92.67} & \textbf{92.08} & \textbf{93.89} & \textbf{92.82} & \textbf{92.07} & \textbf{89.63} & \textbf{92.07} & \textbf{92.08} & \textbf{91.46} \\
& Qwen-plus +CIW +DUG &28.65          & 39.02          & 51.21          & 52.44          & 42.83          & 81.70          & 64.64          & 70.84          & 65.25          & 70.60          \\
\midrule
\multirow{3}{*}{Re-exe}
& GPT-4o +CIW +DUG &18.29          & 7.32           & 10.97          & 7.07           & 10.91          & 25.61          & 10.98          & 14.63          & 6.10           & 14.34          \\
& Deepseek-V3 +CIW +DUG &\textbf{52.84} & \textbf{26.83} & \textbf{34.15} & \textbf{42.68} & \textbf{39.13} & \textbf{64.63} & \textbf{39.64} & \textbf{37.19} & \textbf{37.80} & \textbf{44.82} \\
& Qwen-plus +CIW +DUG &0.00           & 6.71           & 10.37          & 4.88           & 5.34           & 24.39          & 13.42          & 7.93           & 4.88           & 12.66          \\
\midrule
\multirow{3}{*}{ES}
& GPT-4o +CIW +DUG &34.43          & 30.84          & 31.31          & 30.41          & 31.75          & 37.83          & 31.27          & 31.08          & 29.72          & 32.48          \\
& Deepseek-V3 +CIW +DUG &\textbf{42.26} & \textbf{35.45} & \textbf{34.77} & \textbf{33.59} & \textbf{36.52} & \textbf{41.92} & \textbf{34.74} & \textbf{35.99} & \textbf{33.42} & \textbf{36.52} \\
& Qwen-plus +CIW +DUG &10.46          & 15.27          & 17.54          & 15.55          & 14.71          & 29.03          & 22.33          & 21.77          & 22.04          & 23.79 \\        
\bottomrule
\end{tabular}}%
}
\label{human-eval dfg}
\vspace{-2ex}
\end{table*}

In summary, although LLM-driven decompilation approach still lags behind developed decompilers in certain aspects, our work demonstrates unique advantages across several key dimensions. We hope that the LLM-driven decompilation paradigm will facilitate the integration of rules and knowledge from {traditional decompilers} with LLMs, thereby advancing the development of decompilation tools and making them more suitable for convenient deployment and application in complex scenarios.




\subsubsection{The Impact of Compilation Optimization Levels}
From the results across all four tables, we observe a clear trend across all models: decompilation performance generally degrades as the compiler optimization level increases (O0$\to$O1$\to$O2$\to$O3). For example, on the Exebench 64-bit under the Re-exe metric, the average performance at O0 is 48.07\%, O1 is 34.24\%, O2 is 30.40\%, and O3 is 29.18\%. This decline is primarily due to the increasingly aggressive optimizations applied at higher levels (e.g., O2 and O3), such as expression rewriting, variable merging, dead code elimination, and control flow simplification. These transformations significantly alter the original function structure and reduce code readability, making it more challenging for models to accurately reconstruct the source code. Therefore, bridging the gap introduced by compiler optimizations and achieving more robust LLM-based decompilation remains an important direction for future work.

\subsection{Ablation Study (RQ3)}
\label{mainablationstudy}
To evaluate the contribution of individual components in our proposed \textbf{CIW}, we conduct an ablation study\footnote{Ablation study on LLMs are usually conducted during inference rather than through retraining, since retraining LLMs for each ablation setting is resource-intensive — a common limitation in the research community.} (also referred to as vertical experiments). Specifically, for a given LLM, we selectively remove either the CFG or the data mapping table (DM) from the input and observe the resulting impact on decompilation performance. The Tables \ref{nocfg ours} and \ref{nodata ours} present the results\footnote{We conduct the corresponding ablation experiments only on samples where the number of CFG blocks is greater than one or where data sections are present.} of the ablation study on our CIMs. 
More results and analyses are shown in Supplementary Content A.2.
From the results, we have two key findings:
(1)Overall, both CFG and DM improve performance in a complementary manner.
Removing CFG causes average degradations of 10.07\% on Re-exe, 3.21\% on ES and 0.93\% on Re-com; specifically, Re-exe drops by 16.93\% on HumanEval, underlining the essential role of control-flow information for decompilation correctness.
Removing DM leads to smaller but consistent degradations of 1.20\% on Re-com, 4.67\% on Re-exe and 0.36\% on ES, indicating that data-level correspondences still enhance variable alignment and execution fidelity.
(2)Models that integrate both CFG and DM demonstrate more robust performance across different compiler optimization levels (O0–O3). In contrast, models with either component removed show greater performance variance and occasional regressions. This indicates that combining structural (CFG) and symbolic (DM) inductive biases enables better generalization to real-world, compiler-transformed binaries.

\begin{figure*}
  \centering
  \includegraphics[height=1.4in,width=7in]{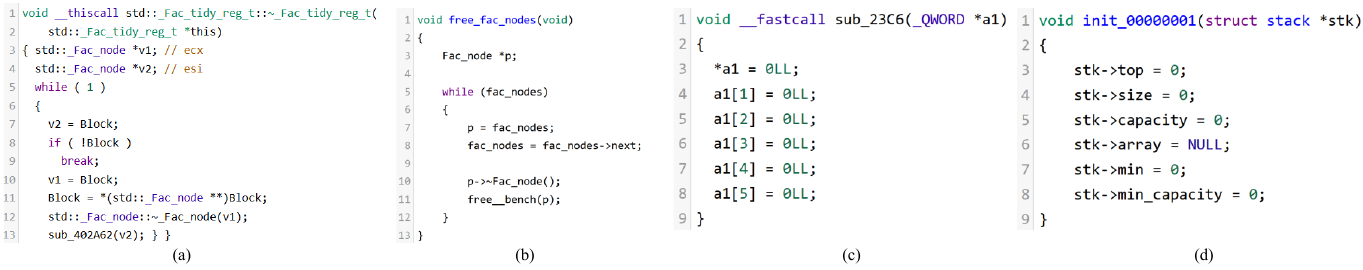}
    \vspace{-3ex}
\caption{{(a) shows the decompilation result of function 0x4029BB in \textit{Clever Bird.exe} produced by Hex-Rays; (b) presents the corresponding decompilation result generated by CIM-6.7B; (c) shows the decompilation result of function 0x23c6 in \textit{revvm} produced by Hex-Rays; and (d) presents the corresponding decompilation result generated by CIM-6.7B.}}
\label{realworld:case}
  \vspace{-2ex}
\end{figure*}

\subsection{The Necessity of Data Flow (RQ4)}

Data flow has been widely utilized in various binary analysis tasks, such as type inference\cite{Zhu2024TYGRTI}, due to its ability to capture dynamic program behaviors and dependencies. Inspired by that, we investigate the impact of data-flow features on LLM-driven decompilation.


We design a def-use graph extraction pipeline based on Ninja~\cite{martin2012ninja}, which operates directly on disassembled code. In the def-use graph, each node represents an assembly instruction, and edges denote data dependencies: directed links run from definition sites (where registers or memory locations are written) to use sites (where they are read), yielding an explicit data-flow view of the binary. This structure enables subsequent phases to trace data flow trajectories and perform data-flow-guided reasoning tasks (e.g., constant propagation, dead-store elimination) during decompilation.

Aligning with the CFG representation in CIW, we first structure each def-use graph (DUG) as a JSON dictionary containing two fields—``nodes'' (individual assembly instructions) and ``edges'' (a list of directed edges)—and then append this serialized JSON string to the CIW prompt.

Table \ref{human-eval dfg} reports decompilation results of general-purpose LLMs on the HumanEval with the assistance of \textbf{DUG-augmented CIW}. By comparing Table \ref{human-eval zero} and Table \ref{human-eval dfg}, we find that incorporating DUG does not yield significant improvements in decompilation performance. On average, the three models achieve absolute gains of 2.94\% on Rec-com, 0.01\% on ES and absolute decreases of 0.19\% on Rec-exe. Several factors may account for this phenomenon: 
(1) The inherently intertwined nature of control and data flows in assembly code implies that they convey overlapping program semantics. As a result, data flows contribute only a limited orthogonal information into the decompilation process.
(2) The granularity and representation of data flow features may not be optimally aligned with the learning objectives of LLMs. Specifically, the DUG may fail to capture critical contextual information necessary for accurate high-level program reconstruction, or introduce excessive noise that dilutes the model's predictive capability.
(3) The inherent limitations of the models themselves may further hinder their ability to effectively utilize data-flow information. The mechanisms required for general-purpose LLMs to handle complex data dependency patterns may remain under-activated, making it difficult to learn the nuanced relationships between data-flow features and their corresponding high-level constructs.

Beyond the main experimental evaluation, we conduct case studies demonstrations. Due to page limitations, we present these content in the Supplementary Content A.3.



{
\subsection{Inference Efficiency Analysis (RQ5)}
To assess the practical applicability of our CIM, we conduct an efficiency analysis beyond decompilation performance on HumanEval. We evaluate computational costs in terms of inference latency and GPU memory consumption, and compare our CIM with larger LLM baselines. This analysis quantifies the computational requirements and scalability of our approach.
}

{
Specifically, CIM-1.3B requires at least 3GB GPU memory and takes 369s on 2×H100, while CIM-6.7B requires at least 14GB GPU memory and takes 959s. Under our API evaluation setting with approximately 1 Gbps bandwidth, Bailian-API-based DeepSeek-V3 and Qwen-plus take 2147s and 3734s, respectively. 
}

{
The performance and efficiency analyses show that our proposed models achieve competitive or superior performance with substantially lower computational costs compared with larger baseline LLMs.
}

{
\subsection{Real-world Application}
\label{sec:real_world}
To validate the practical effectiveness of our CIM, we randomly select binaries such as \textit{Clever Bird.exe} and \textit{revvm} from reverse engineering challenges in past \textbf{CTF} competitions (XNUCA 2019 and hxp CTF 2021) and perform decompilation on them (as shown in Fig. \ref{realworld:case}). From the decompiled results of functions such as ‘0x4029BB’ in \textit{Clever Bird.exe} and ‘0x23c6’ in \textit{revvm}, it can be observed that our decompilation-specific LLM outperforms Hex-Rays in terms of variable/function renaming, structure representation, and overall code readability.
}

\section{Discussion}


Through different-bit experiments, we find that LLM-based decompilers exhibit promising scalability and generalization in terms of development and maintenance; extending them to new compilers and architectures is essentially an engineering increment driven by data diversity. 

Although our exploratory study has made some progress, several important challenges remain open for future investigation:

(1) The experiments indicate a significant drop in decompilation performance on binaries compiled with higher optimization levels. Understanding the underlying reasons behind this degradation and designing models that are more robust to aggressive compiler optimizations are promising directions for future research.

(2) We currently focus on the single-function in binaries, and in future work, we will further explore the decompilation of entire binaries.

(3) Existing works perform one-time decompilation. In practice, it is more desirable to build a self-correcting decompiler, which can refine its own decompilation results through execution feedback and comparison with the assembly.

{
(4) Mitigating LLM hallucinations is critical, as generated code may contain fabricated or inconsistent logic that reduces the success rate of binary decompilation.
}

{
(5) Applicability to complex binaries. Current work primarily focuses on standard binaries with available program information. Extending LLM-based decompilation to stripped, obfuscated, packed, and protected binaries remains an open challenge due to the loss of program semantics and increased analysis complexity.
}


\section{Conclusion}
\label{conclusion}

Despite recent advances, end-to-end decompilation with LLMs still suffers from limited control flow awareness, weak data context modeling, and high computational costs. This paper addresses these challenges by introducing a structure-aware and data-informed decompilation scheme that integrates CFGs and data mapping tables into LLM inputs. We build a dedicated dataset and propose CIM, a lightweight domain-specific LLM, and demonstrate that it achieves state-of-the-art performance on multi-metrics
while significantly reducing computational costs to enable efficient and locally deployable inference.



\section{Data Availability Statement}
The complete training and evaluation code and dataset, along with the optional Supplementary Content of this paper, are archived on Zenodo at
\url{https://doi.org/10.5281/zenodo.19351453}, and also on Github at \url{https://github.com/LiuPeiP-CS/CodeInverter}.

\begin{acks}
Supported by Zhongguancun Laboratory. We thank the anonymous reviewers for their insightful comments and constructive suggestions.
\end{acks}
\balance
  \bibliographystyle{ACM-Reference-Format}
  \bibliography{samples/refers}

\end{document}